\pdfoutput=1
\documentclass[11pt]{article}

\usepackage{acl}

\usepackage{times}
\usepackage{latexsym}
\usepackage{stmaryrd}
\usepackage{latexsym}
\usepackage{graphicx}
\usepackage{amsmath}
\usepackage{cleveref}
\usepackage{float}
\usepackage{booktabs}
\usepackage{arydshln}
\usepackage{hyperref}
\usepackage{color,soul}
\usepackage[T1]{fontenc}
\usepackage[utf8]{inputenc}
\usepackage{amsfonts,amssymb}
\usepackage{microtype}

\usepackage{inconsolata}

\usepackage{amsmath}
\usepackage{bm}
\usepackage{graphicx}
\usepackage{booktabs}
\usepackage{multirow}
\usepackage{xcolor}
\usepackage{bbding}
\usepackage{comment}
\usepackage{multicol}

\newcommand{\model}{\textsc{CM-TTS}}
\title{\model{}: Enhancing Real Time Text-to-Speech Synthesis Efficiency through Weighted Samplers and Consistency Models}

\author{
	Xiang Li$^1$, Fan Bu$^1$, Ambuj Mehrish$^2$, Yingting Li$^1$, Jiale Han$^1$, \\
 \textbf{Bo Cheng}$^1$, \textbf{Soujanya Poria}$^2$
	\\
 	\textsuperscript{1}State Key Laboratory of Networking and Switching Technology,\\Beijing University of Posts and Telecommunications\\
	\textsuperscript{2}Singapore University of Technology and Design\\
	\texttt{\{lixiang2022,bufan,cindyyting,hanjl,chengbo\}@bupt.edu.cn} \\\texttt{\{ambuj\_mehrish,sporia\}@sutd.edu.sg}
 }

\begin{document}
	\maketitle
\begin{abstract}
Neural Text-to-Speech (TTS) systems find broad applications in voice assistants, e-learning, and audiobook creation. The pursuit of modern models, like Diffusion Models (DMs), holds promise for achieving high-fidelity, real-time speech synthesis. Yet, the efficiency of multi-step sampling in Diffusion Models presents challenges. Efforts have been made to integrate GANs with DMs, speeding up inference by approximating denoising distributions, but this introduces issues with model convergence due to adversarial training. To overcome this, we introduce \model{}, a novel architecture grounded in consistency models (CMs). Drawing inspiration from continuous-time diffusion models, \model{} achieves top-quality speech synthesis in fewer steps without adversarial training or pre-trained model dependencies. We further design weighted samplers to incorporate different sampling positions into model training with dynamic probabilities, ensuring unbiased learning throughout the entire training process. We present a real-time mel-spectrogram generation consistency model, validated through comprehensive evaluations. Experimental results underscore \model{}'s superiority over existing single-step speech synthesis systems, representing a significant advancement in the field\footnote{Code and generated samples are available at: \url{https://github.com/XiangLi2022/CM-TTS}.}.
\end{abstract}
	
\section{Introduction}
The modern Neural Text-to-Speech (TTS) system \citep{mehrish2023review,shen2018natural, ren2020fastspeech, liu2022diffgan} stands out for its exceptional naturalness and efficiency, proving versatile in human-computer interaction and content generation scenarios like real-time voice broadcasting and speech content creation. Comprising three integral modules, the system involves a text encoder collaborating with a conditioning feature predictor, followed by an acoustic model transforming conditioning features into speech features, and a vocoder converting synthesized features into audible speech. This intricate process ensures efficient synthesis of human-like speech.

From a formulation perspective, TTS architecture aligns with autoregressive (AR) (\citealp{oord2016wavenet}; \citealp{amodei2016deep}; \citealp{wang2017tacotron}; \citealp{shen2018natural}) and non-autoregressive (NAR) (\citealp{ren2019fastspeech}; \citealp{ren2020fastspeech}) models. AR frameworks, using RNN models with attention mechanisms, generate spectrograms sequentially, ensuring stable synthesis but suffering from accumulated prediction errors and slower inference speeds. Conversely, NAR models, often based on transformer architecture \citep{vaswani2017attention}, employ parallel feed-forward networks for simultaneous mel-spectrogram generation, reducing computational complexity and enabling real-time applications. Various generative models, including Generative Adversarial Networks (GANs) (\citealp{kumar2019melgan}; \citealp{kong2020hifi}; \citealp{donahue2020end}), Flow (\citealp{kim2018flowavenet,kim2020glow,shih2021rad,valle2020flowtron})-based models, and hybrid approaches like Flow with GAN (\citealp{cong2021glow}), contribute to high-fidelity, real-time speech synthesis.
Diffusion Models (DMs) are advanced generative models, excelling in image generation (\citealp{ho2020denoising,kumar2019melgan,song2020score,rombach2021highresolution}), molecular design \cite{DBLP:conf/nips/YouLYPL18, DBLP:journals/corr/Gomez-Bombarelli16, thomas2023integrating}, and speech synthesis \cite{pmlr-v162-kim22d, DBLP:journals/corr/abs-2205-15370, popov2021grad}. Employing a forward diffusion process with noise addition and a parameterized reverse iterative denoising process, DMs efficiently capture high-dimensional data distributions. Despite their exceptional performance, the efficiency of their multi-step iterative sampling is hindered by Markov chain limitations. To address these challenges, \citet{ye2023comospeech} propose a TTS architecture based on consistency models \citep{song2023consistency}. This architecture achieves high audio quality through a single diffusion step, applying a consistency constraint to distill a model from a well-designed diffusion-based teacher model. However, a drawback is the method's reliance on distillation from a teacher model, introducing complexity into the training pipeline. Importantly, their proposed TTS architecture is trained on the single-speaker LJSpeech dataset \citep{ljspeech17}, limiting its suitability for multi-speaker speech generation. This constraint should be considered in applications where broader speaker diversity is essential.
% Diffusion Models (DMs) (\citealp{ho2020denoising,kumar2019melgan,song2020score}) have emerged as a cutting-edge category of deep generative models, showcasing remarkable performance across diverse domains such as image generation, molecular design, and speech synthesis (\citealp{xie2022measurement}; \citealp{yang2023diffsound}; \citealp{levkovitch2022zero}). Typically, these models involve a forward diffusion process that smoothly perturbs data through the addition of noise, coupled with a parameterized reverse iterative denoising process to reconstruct the original data. While the design of DMs excels in effectively capturing the distribution of high-dimensional data, their multi-step iterative sampling methods encounter challenges related to reduced efficiency, primarily stemming from the limitations of Markov chains. To address the multi-step iterative sampling process authors in \citep{ye2023comospeech} propose TTS architecture based on consistency models \citep{song2023consistency}. The proposed architecture achieves high audio quality through a single diffusion sampling step by applying a consistency constraint to distill a consistency model from a well-designed diffusion-based teacher model. However,  one limitation of the proposed method is that it requires distillation from a teacher model for better performance, making the pipeline of constructing a speech synthesis system more complicated. Furthermore, the proposed TTS architecture is trained using single-speaker LJSpeech dataset limiting its scope for multi speaker speech generation.

The integration of GANs into DMs for TTS synthesis (\citealp{liu2022diffgan}) has proven effective in minimizing the number of sampling steps during the speech synthesis process. However, this improvement comes at the cost of hindered model convergence due to the additional training required for the discriminator. Some approaches enhance synthesis performance with fewer inference steps by incorporating a shallow diffusion mechanism (\citealp{liu2022diffgan}). Nonetheless, the introduction of an additional pre-trained model adds complexity to the overall architecture. 

We present a novel TTS architecture, \model{}, addressing current limitations without relying on a teacher model for distillation. Drawing inspiration from continuous-time diffusion and consistency models, our approach frames speech synthesis as a generative consistency procedure, achieving superior quality in a single step. \model{} eliminates the need for adversarial training \citep{liu2022diffgan} or auxiliary pre-trained models \citep{ye2023comospeech}. We enhance model training efficacy with weighted samplers, mitigating sampling biases. \model{} maintains traditional diffusion-based TTS benefits and introduces a few-step iterative generation, balancing synthesis efficiency and quality. Experimental results confirm \model{} outperforms other single-step speech synthesis systems in quality and efficiency, presenting a significant advancement in TTS architecture. Our key contributions can be summarized as follows:
	
% To address the limitations of existing methods, we introduce a novel TTS architecture based on a consistency model without reliance on distillation from a teacher model. Drawing inspiration from the theory of (continuous-time) diffusion models (\citealp{song2020score}) and consistency models (\citealp{song2023consistency}), we model the speech synthesis process as a consistency generation process. Remarkably, CM-TTS achieves high-quality speech synthesis in a single step without relying on adversarial training or additional pre-trained model assistance. To further enhance the model training process, we introduce Weighted Samplers, mitigating biases introduced by sampling and thereby improving overall model efficacy. CM-TTS also preserves the advantages of traditional Diffusion-based TTS, the model facilitates few-step iterative generation, enabling a balanced trade-off between synthesis efficiency and synthesis quality. Experimental results demonstrate that CM-TTS outperforms other known single-step speech synthesis systems known to us in both synthesis quality and efficiency. \hl{(add zero shot)} Our key contributions can be summarized as follows:\\

\begin{itemize}
  \item We present a consistency model-based architecture for generating a mel-spectrogram designed to meet the demands of real-time speech synthesis with its efficient few-step iterative generation process.

  \item Moreover, CM-TTS can also synthesize speech in a single step, eliminating the need for adversarial training and pre-trained model dependencies. 
  % Additionally, the integration of the variance adapter with the consistency model decoder enables multi-speaker speech generation.

  \item We enhance the model training process by introducing weighted samplers, which adjust weights associated with different sampling points. This refinement mitigates biases introduced during model training due to the inherent randomness of the sampling process.
  
  \item Qualitative and quantitative experiments covering 12 metrics demonstrate the effectiveness and efficiency of our model in both fully supervised and zero-shot settings.
\end{itemize}

\section{Related Work}
\paragraph{Non-Autoregressive Generative Models}
% Non-autoregressive generative models (NAR) achieve parallelized output generation, significantly improving generation speed and suitability for real-time applications. Their efficiency in processing long sequences, attributed to the lack of dependence on previous generation results, has led to widespread applications in domains such as image generation and speech synthesis. GAN networks, leveraging their ingenious adversarial ideas and promising architectures, have been widely applied in non-autoregressive speech synthesis. \citet{donahue2020end}, based on adversarial training principles, utilized a differentiable alignment scheme and designed a generator to achieve end-to-end speech synthesis. \citet{kim2021conditional} introduced the adversarial training process into Variational Autoencoders (VAE) (\citealp{kingma2019introduction}), enabling single-stage training and sampling to enhance the model's expressive power in speech generation. However, GANs exhibit notable instability during the training process, primarily due to the non-overlapping distributions between input and generated data. To address this and enable the model to capture a more representative data distribution, we embraced the principles of the Diffusion Model, resulting in the design of CM-TTS. This involves forward diffusion and reverse denoising, facilitating model training and mel-spectrogram generation.

Non-autoregressive generative models (NAR) excel in swiftly generating output, making them ideal for real-time applications. Their efficiency, derived from parallelized output generation and lack of dependence on previous results, finds applications in diverse domains like image generation and speech synthesis. GAN networks have been applied in non-autoregressive speech synthesis. \citet{donahue2020end} employ adversarial training and a differentiable alignment scheme for end-to-end speech synthesis. Additionally, \citet{kim2021conditional} integrate adversarial training into Variational Autoencoders (VAE)(\citep{kingma2019introduction}), enhancing expressive power in speech generation. However, GANs face training instability due to non-overlapping distributions between input and generated data. To address this, \model{} incorporates Diffusion Model principles for improved model training and mel-spectrogram generation.

\paragraph{Diffusion Models (DMs)}

DMs provide robust frameworks for learning complex high-dimensional data distributions through continuous-time diffusion processes. After surpassing GANs (\citealp{dhariwal2021diffusion}) in image synthesis, DMs have shown promise in speech synthesis. \citet{jeong2021diff} utilize a denoising diffusion framework for efficient speech synthesis, transforming noise signals into mel-spectrograms. 
While DMs excel in data distribution modeling, they may require numerous network function evaluations (NFEs) during sampling. Combining diffusion modeling with traditional generative models enhances efficiency. Diff-GAN (\citealp{liu2022diffgan}) adopts an adversarially trained model for expressive denoising distribution approximation. \citet{yang2023diffsound} use VQ-VAE (\citealp{van2017neural}) to transfer text features to mel-spectrograms, reducing diffusion model computational complexity.

\section{Background: Consistency Models}
The diffusion model is distinguished by a sequential application of Gaussian noise to a target dataset, followed by a subsequent reverse denoising process \citep{ho2020denoising}. This iterative methodology is designed to generate samples from an initially noisy state, effectively capturing the intrinsic structure of the data. Consider the sequence of noisy data $\left\{ x \right\}_{t\in[0, T]}$, where $p_{0}(\mathbf{x}) \equiv p_{\text{data}}(\mathbf{x})$, $p_{T}(\mathbf{x})$ approximates a Gaussian distribution, and $T$ represents the time constant. The diffusion process can be mathematically expressed as a stochastic process using following stochastic differential equation (SDE).
\begin{equation}
    \mathbf{x}_t=\bm{\mu}(\mathbf{x}_t,t)\textrm{dt}+\sigma(t)\textrm{d}{\mathbf{w}_t}
    \label{sde}
\end{equation}
where $ t\in[0,T]$, is the index for forward diffusion time steps. Here, $ \bm{\mu}(.,.) $ and $ \sigma(.) $ correspond to the drift and diffusion coefficients, and $ \left\{ w_{t} \right\}_{t\in[0, T]} $ denotes the standard Brownian motion.

A fundamental characteristic of the SDE lies in its inherent possession of a well-defined reverse process, manifested in the form of a probability flow ODE (\citealp{song2020score,karras2022elucidating}). Consequently, the trajectories sampled at time $t$ follow a distribution governed by $p_{t}(\mathbf{x}_t)$:
\begin{equation}
    \textrm{d}{\mathbf{x}_t} =\left[ \mu(\mathbf{x}_t,t)-\frac{1}{2}\sigma(t)^{2}\nabla \log\,p_{t}(\mathbf{x}_t) \right]\;\textrm{d}{t}
    \label{pdode}
\end{equation}
$\nabla \log p_{t}(\mathbf{x}_t)$ represents the score function, a key element in score-based generative models \cite{song2020score}. The forward step induces a shift in the sample away from the data distribution, dependent on the noise level. Conversely, a backward step guides the sample closer to the expected data distribution. The probability flow ODE (referenced as Eq.~\ref{pdode}) for sample generation utilizes the score function $\nabla \log p_{t}(\mathbf{x}_t)$. Obtaining the score function involves minimizing the denoising error $||f(x_{t},t)-x||^{2}$ \citep{karras2022elucidating}, where $f(x_{t},t)$ is the denoiser function refining the sample $x_{t}$ at step $t$.
% $\nabla \log p_{t}(\mathbf{x}_t)$ denotes the score function linked to $p{t}(\mathbf{x})$, a defining feature of score-based generative models widely discussed in the literature \citep{}. In the context of a probability flow ODE, the forward step induces a shift in the sample away from the data distribution. The extent of this shift correlates with the noise level. Conversely, a backward step steers the sample closer to the expected data distribution. Utilizing the score function $\nabla \log p_{t}(\mathbf{x}_t)$ allows the probability flow ODE (referenced as Eq.~\ref{pdode}) for sample generation. Obtaining the score function involves minimizing the denoising error $||f(x_{t},t)-x||^{2}$ \citep{karras2022elucidating}, where $f(x_{t},t)$ signifies the denoiser function adept at refining the sample $x_{t}$ at step $t$. 
\begin{equation}
    \nabla \log p_{t}(\mathbf{x}_t) = \frac{(f(x_{t},t)-x_{t})}{\sigma(t)^{2}}
\end{equation}

Probability flow ODEs sampling follows a two-step approach: first, samples are drawn from a noise distribution, and then, a denoising process is applied using a numerical ODE solver, like Euler or Heun \citep{song2020score,song2023consistency}. However, the sampling process from the ODE solver requires a substantial number of iterations, leading to the drawback of slow inference speed. To further accelerate the sampling \citet{song2023consistency} proposed a consistency property for the diffusion model  with the following condition for any time step $t$ and $t^{'}$of a solution trajectory.
\begin{equation}
\begin{split}
      f(x_{t},0) =& f(x_{t^{'}},t^{'})\\
    f(x_{t},0) =& x_{0}
\end{split}
\label{property}
\end{equation}

Given the aforementioned condition, one-step sampling $f(x_{T}, T)$ becomes viable, as each point along the sampling trajectory of the ODE is directly associated with the origin $p_{0}(x)$. For a more in-depth discussion, refer to \citet{song2023consistency}. The consistency model is categorized into two types: consistency training or distillation from a pre-trained diffusion-based teacher model. The distillation-based approach relies on the teacher model, adding intricacy to the construction pipeline of the speech synthesis system. In this work, we opt for consistency training of the consistency model.

\section{\model{}}
\begin{figure*}
    \centering
    \includegraphics[height=0.8\columnwidth]{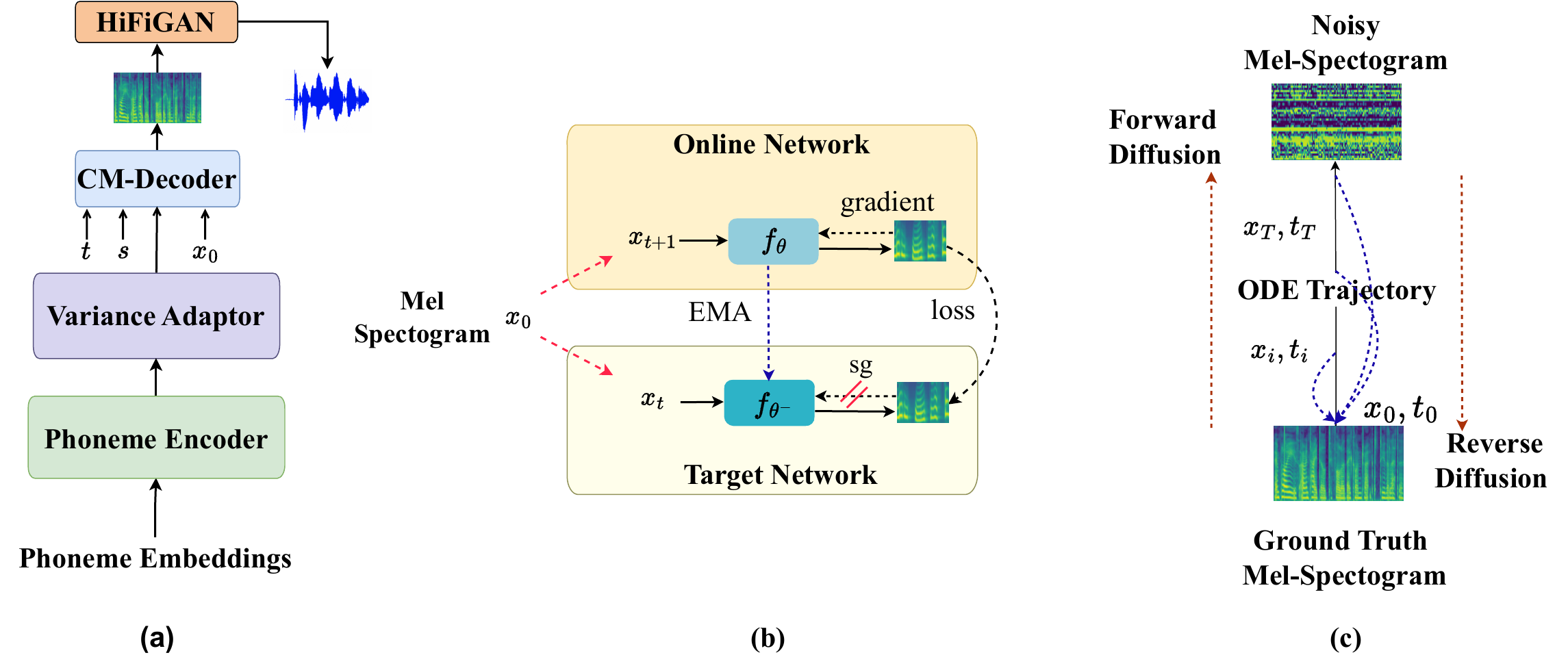}
    \caption{(a) \model{} architecture. (b) Decoder training scheme, where $f_{\theta}$ is parameterized to satisfy consistency constrain disucssed in Eq.~\ref{property}. (c) ODE trajectory during training.}
    \label{architecture}
\end{figure*}

Diffusion models, known for their high-quality outputs, often struggle with real-time demands in TTS systems due to slow sampling. Existing attempts, like Diff-GAN (\citealp{liu2022diffgan}), often rely on additional adversarial training or pre-trained models for efficiency and accuracy. In this section, we discuss the architecture of \model{}.

\subsection{Model Overview}
As shown in Figure~\ref{architecture}, the \model{} consists of four key components: 1) Phoneme encoder for processing text; 2) Variance adaptor predicting pitch, duration, and energy features; 3) the CM-Decoder for mel-spectrogram generation; and 4) Vocoder, using HiFi-GAN (\citealp{kong2020hifi}), to convert mel-spectrograms into time-domain waveforms.
\subsection{Phoneme Encoder and Variance Adaptor}
The phoneme encoder, incorporating multiple Transformer blocks \citep{ren2019fastspeech,ren2020fastspeech}, adapts the feed-forward network to effectively capture local dependencies within the phoneme sequence. The variance adaptor aligns with FastSpeech2's design, including pitch, energy, and duration prediction modules, each following a consistent model structure with several convolutional blocks. To facilitate training, ground-truth duration, energy, and pitch serve as learning targets, computed using Mean Squared Error (MSE) loss ($\mathcal{L}_{\text{duration}}$, $\mathcal{L}_{\text{pitch}}$, and $\mathcal{L}_{\text{energy}}$). In the training phase, the ground-truth duration expands the hidden sequence from the phoneme encoder to yield a frame-level hidden sequence, followed by the integration of ground-truth pitch information. During inference, the corresponding predicted duration and pitch values are utilized.
\subsection{Consistency Models}
\label{cm}
To establish the divisions within the time horizon $[\epsilon, T_{\text{max}}]$, the interval is segmented into $N-1$ sub-intervals, delineated by boundaries $t_{1} = \epsilon < t_{2} < \ldots < t_{N} = T_{max}$. As recommended by \citet{karras2022elucidating} to mitigate numerical instability, a small positive value is set for $\epsilon$. Similar to \citet{karras2022elucidating}, in this work we use $T_{max}=80$ and $\epsilon=0.002$. The mel-spectrogram is denoted as $\mathbf{x}$, where $\mathbf{x}_{0}$ signifies the initial mel-spectrogram devoid of any added noise.

The fundamental concept introduced in \citet{song2023consistency} to formulate the consistency model $f_{\theta}$ involves learning a consistency function from data by enforcing the self-consistency property defined in Eq.~\ref{property}. In order to ensure $f_{\theta}(x_{0},\epsilon) = \mathbf{x_{0}}$, the consistency model $f_{\theta}$ is parameterized as follows:
\begin{equation}
f_{\theta}(\mathbf{x},t)=c_{skip}(t)\mathbf{x}+c_{out}(t)F_{\theta}(\mathbf{x},t)
\end{equation}
Here, $c_{skip}$ and $c_{out}$ are differentiable functions with $c_{skip}(\epsilon)=1$ and $c_{out}(\epsilon)=0$, respectively. The term $F_{\theta}(\mathbf{x},t)$ represents a neural network. To enforce the self-consistency property, a target model $\theta^{-}$ is concurrently maintained with the online network $\theta$. The weight of the target network $\theta^{-}$ is updated using the exponential moving average (EMA) of parameters $\theta$ intended for learning \citep{grill2020bootstrap}, specifically, 
\begin{equation}
\bm{\theta^{-}}\leftarrow\textrm{stopgrad}(\mu\bm{\theta^{-}}+(1-\mu)\bm{\theta}).
\label{decay}
\end{equation}

The consistency loss $\mathcal{L}_{CT}^{N}(\bm{\theta,\theta^{-}})$ is defined as:
\begin{equation}
%\small
\sum_{n\geq1}{\mathbb{E}[\lambda(t_{n})d(\bm{f_{\theta}}(\mathbf{x}_{t+1}), \bm{f_{\theta^{-}}}(\mathbf{x}_{t}))]}
\end{equation}
Here, $d(\cdot,\cdot)$ denotes a chosen metric function for measuring the distance between two samples, such as the squared $l_{2}$ distance $d(x,y) = ||x-y||^{2}_{2}$. The values $\mathbf{x}_{t+1}$ and $\mathbf{x}_{t}$ are obtained by sampling two points along the trajectory of the probability flow ODE using a forward diffusion process, starting with mel-spectrograms of the training data $\mathbf{x}_0\sim \mathcal{D}(dataset)$:
\begin{equation}
\begin{split}
\mathbf{x}_{t+1}=&\mathbf{x}_0+ t_{n+1}\mathbf{z}\\
\mathbf{x}_{t}=&\mathbf{x}_0+ t_{n}\mathbf{z}
\end{split}
\label{diffprocess}
\end{equation}
where $\mathbf{z}\sim \mathcal{N}(\bm{0,I})$ and step $t_{n}$ is obtained as follows:
\begin{equation}
t_n=\left[T_{\max }{ }^{\frac{1}{p}}+\frac{n-1}{N-1}\left(\epsilon{ }^{\frac{1}{p}}-T_{\max }{ }^{\frac{1}{p}}\right)\right]^p 
\label{sample}
\end{equation}
where $\textit{N}$ denotes the sub-intervals, $n$ is sampled from the interval $[{1,N-1}]$ using different weighted sampling strategies (Section \ref{sec:Weighted Sampler}), and value of $p = 7$ following \citet{karras2022elucidating}.

Similar to DiffGAN-TTS \citep{liu2022diffgan}, the architecture of $F_{\theta}(\mathbf{x}, t)$ in \model{} embraces a non-causal WaveNet structure \citep{oord2016wavenet}. The difference lies in their approach to sampling $t$. In \model{}, two decoders, denoted as $f_{\theta}$ and $f_{\theta}^{-}$, with identical architectures serve as the online and target networks, respectively. The diffusion process in \model{} is characterized by Eq.~\ref{diffprocess}, whereas DiffGAN-TTS employs the creation of a parameter-free $T$-step Markov chain \citep{liu2022diffgan}.
\subsubsection{Training and Loss}
% Figure A(c) elucidates the training procedure of CM-TTS. Starting with mel-spectrograms of the training data, denoted as $\mathbf{x}_0$, $\mathbf{x}_0\sim \mathcal{D}(dataset)$. we initiate a forward diffusion process resulting in two sampling points along the trajectory of the Ordinary Differential Equation (ODE), designated as $\mathbf{x}_{t+1}$ and $\mathbf{x}_{t}$.
% \begin{eqnarray*}
% \mathbf{x}_{t+1}=\mathbf{x}_0+ t_{n+1}\mathbf{z}\\
% \mathbf{x}_{t}=\mathbf{x}_0+ t_{n}\mathbf{z}
% \end{eqnarray*}
% \hl{$\textit{N}(\cdot)$ means the step schedule}, $n\sim \mathcal{U} \llbracket{1,N(k)-1}\rrbracket$ and $\mathbf{z}\sim \mathcal{N}(\bm{0,I})$
Following the training procedure established in \citet{grill2020bootstrap}, we designate the two decoders shown in Figure~\ref{architecture} as the online $\bm{f_{\theta}}$ and target $\bm{f_{\theta^{-}}}$. Leveraging the states $\mathbf{x}_{t+1}$ and $\mathbf{x}_{t}$, we derive corresponding mel predictions, expressed as $\bm{f_{\theta}}(\mathbf{x}_0 + t_{n+1}\mathbf{z})$ and $\bm{f_{\theta^{-}}}(\mathbf{x}_0 + t_{n}\mathbf{z})$, through the online and target networks, respectively. The online component undergoes gradient updates via the computation of MSE loss between these prediction pairs. Simultaneously, the gradients of the target network are updated through EMA, as discussed in section~\ref{cm}.
% Aligning with the convention used in BYOL (\citealp{grill2020bootstrap}), we denote the two decoders in the figure as the online $\bm{f_{\theta}}$ and target $\bm{f_{\theta^{-}}}$. By employing $\mathbf{x}_{t+1}$ and $\mathbf{x}_{t}$, we derive corresponding mel predictions denoted as $\bm{f_{\theta}}(\mathbf{x}_0+ t_{n+1}\mathbf{z})$ and $\bm{f_{\theta^{-}}}(\mathbf{x}_0+ t_{n}\mathbf{z})$ through the online and target networks, respectively. In the online component, the network's gradient update is achieved by calculating the Mean Squared Error (MSE) loss between these two predictions. Simultaneously, the target network gradients are updated using EMA  as discussed in previous section.
% \begin{equation*}
% \bm{\theta^{-}}\leftarrow\textrm{stopgrad}(\mu\bm{\theta^{-}}+(1-\mu)\bm{\theta})
% \end{equation*}

% Throughout this process, the weight $\mu$ undergoes exponential decay, gradually diminishing the influence of the most recent seitan on the target gradient update. Therefore, the consistency training objective can be denoted as $\mathcal{L}_{CT}^{N}(\bm{\theta,\theta^{-}})$, is defineded as follow, $d(.,.)$ is a metric function. 
% \begin{equation*}
% \sum_{n\geq1}{\mathbb{E}[\lambda(t_{n})d(\bm{f_{\theta}}(\mathbf{x}_0+t_{n+1}\mathbf{z}),\bm{f_{\theta^{-}}}(\mathbf{x}_0+t_{n}\mathbf{z}))]}
% \end{equation*}
During training, the online and target networks engage in an iterative interplay, facilitating mutual learning and crucially contributing to model stability. The mel reconstruction loss $\mathcal{L}_{mel}$ is determined by computing the Mean Absolute Error (MAE) between the ground truth and the generated mel-spectrogram. Finally, $\mathcal{L}_{recon}$ can be expressed as follows:
% Over the entire training duration, the online and target networks engage in mutual interaction and learning, and it is precisely this dual mechanism that ensures the stability of the model training. Furthermore, the overall loss function of CM-TTS encompasses an acoustic reconstruction loss, computed in accordance with the methodology presented in FastSpeech2.
\begin{equation}
%\small
\begin{aligned}
\mathcal{L}_{recon}= & \mathcal{L}_{mel}(\mathbf{x_{0}},\mathbf{\hat{x}_0})+\lambda_{d}\mathcal{L}_{duration}(\mathbf{d},\mathbf{\hat{d}})+\\
&\lambda_{p}\mathcal{L}_{pitch}(\mathbf{p},\mathbf{\hat{p}})+\lambda_{e}\mathcal{L}_{energy}(\mathbf{e},\mathbf{\hat{e}})
\end{aligned}
\label{equation_recon}
\end{equation}
Here, $\mathbf{d}$, $\mathbf{p}$, and $\mathbf{e}$ denote the ground truth duration, pitch, and energy, respectively, while $\mathbf{\hat{d}}$, $\mathbf{\hat{p}}$, and $\mathbf{\hat{e}}$ represent the predicted values. The weights assigned to each loss component are denoted by $\lambda_{d}$, $\lambda_{p}$, and $\lambda_{e}$. For this study, we maintain uniform loss weights set at $0.1$. The optimization objective for training the \model{} involves minimizing the following composite loss function.
\begin{equation}
\mathcal{L}_{CM-TTS}=\mathcal{L}_{CT}^{N}(\bm{\theta,\theta^{-}})+\mathcal{L}_{recon}
\end{equation}

\begin{figure}
    \centering
    \includegraphics[scale=0.6]{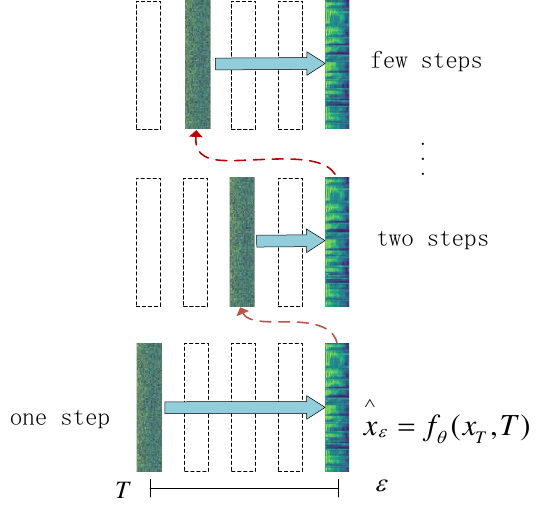}
    \caption{Single-step and multi-step inference utilizing the \model{}. For multi-step generation, process of alternating denoising and noise injection steps is executed iteratively until the desired number of steps is achieved. }
    \label{inference}
\end{figure}

During single-step generation in inference, a single forward pass through $f_{\theta}$ is undertaken. Conversely, multi-step generation is achievable by alternating denoising and noise injection steps, enhancing the quality, as depicted in Figure~\ref{inference}.

\begin{table*}[t]
\centering
\renewcommand\tabcolsep{3.6pt}
%\small
\scalebox{0.7}
{
    \begin{tabular}{lcccccccccccc}
    %{lp{1.3cm\textwidth}p{1.1cm\textwidth}ccccccp{1.2cm\textwidth}p{1cm\textwidth}p{1.2cm\textwidth}c} %{lccccccccccc}
    \toprule 
    Model& FFE$\downarrow$ & S.Cos$\uparrow$ & mfccFID$\downarrow$ & melFID$\downarrow$ & mfccRecall$\uparrow$ & MCD$\downarrow$ & SSIM$\uparrow$ & mfccCOS$\uparrow$ & F0$\downarrow$ & RTF$\downarrow$ & WER$\downarrow$ & MOS$\uparrow$ \\
    \midrule
    Reference&-&-&-&1.46e-11&0.6428&-&-&-&-&-&0.0300&-\\
    Reference (voc.)&0.1427&0.9424&31.98&3.48&0.5644&4.57&0.8132&0.8457&89.21&&0.0412&4.5826(±0.1147) \\ \midrule
    FastSpeech2&0.3503&0.8236&43.42&8.82&0.3554&5.89&0.4537&0.7565&119.21&\textbf{0.02}&0.0677&3.6821(±0.1762) \\
    VITS&0.3509&0.8154&428.91&15.40&\textbf{0.5141}&6.96&0.4411&0.7418&117.99 &0.23&\textbf{0.0451}&3.6717(±0.0123) \\
    DiffSpeech&\textbf{0.3343}&0.7400&76.01&11.55&0.5096&7.25&0.3421&0.6445&119.98 &9.19&0.5708&2.9157(±0.0594) \\ \midrule
    DiffGAN-TTS(T=1)&0.3489&0.8284&97.65&20.01&0.3560&5.98&0.4589&0.7537&118.47&\textbf{0.02}&0.0809&3.4476(±0.1038)\\
    DiffGAN-TTS(T=2)&0.3411&0.8333&38.64&7.79&0.3974&5.94&0.4610&0.7581&\textbf{117.19}&0.03&0.0827&3.6173(±0.1433)\\
    DiffGAN-TTS(T=4)&0.3465&0.8358&\textbf{37.11}&\textbf{6.58}&0.3662&5.94&0.4614&0.7571&120.10&0.04&0.0751&3.6143(±0.1186)\\ \midrule
    \model{}(T=1)&0.3387&0.8396&39.17&7.58&0.3946&5.91&0.4772&0.7599&119.29&\textbf{0.02}&0.0688&\textbf{3.9618(±0.0186)}\\
    \model{}(T=2)&0.3383&\textbf{0.8401}&38.79&7.34&0.3972&\textbf{5.90}&0.4780&0.7598&120.01&0.03&0.0680&3.8947(±0.0262)\\
    \model{}(T=4)&0.3385&0.8399&38.78&7.34&0.3976&\textbf{5.90}&\textbf{0.4783}&\textbf{0.7599}&119.23&0.07&0.0696&3.8623(±0.0311)\\
    \bottomrule
    \end{tabular}
}
\caption{Objective and subject evaluation: Comparison with baselines on VCTK dataset.}
\label{baselines}
\end{table*}

\subsubsection{Weighted Sampler}
\label{sec:Weighted Sampler}
The training procedure relies on sampling the time step $t_{n}$ as defined in Eq.~\ref{sample}. Consequently, to investigate the impact of sampling various positions ($t_{n}$) along the ODE trajectory, we employ three distinct weighted sampling strategies. Each strategy governs the probabilities associated with selecting the step $t_{n}$ throughout the training, thereby allowing for an in-depth examination of the effects arising from different sampling positions.

In the forward diffusion process during training, the variable $n$ denotes the index of a sampling point, where $n\in[1, N-1]$, and is used in Eq.~\ref{sample} for computing $t_{n}$. We introduce $c_{n}$ as the weight assigned to the current index $n$ by the sampler, $s_{n}$ the probability of selecting index $n$ is given by $ s_{n}=\frac{c_{n}}{\sum_{i=1}^{N-1}{c_{n}}}$. The three sampler designs are outlined as follows:

% Throughout the CM-TTS training process, our observations revealed variations in the contribution of sampling points at different positions along the ODE trajectory to the final model performance. Consequently, in pursuit of optimizing the model training process, we introduced three distinct types of samplers, referred to as Weighted Sampler. These samplers are designed to selectively choose points along the trajectory with varying probabilities. The specific design is detailed below.

% When sampling during the forward diffusion process for training the CM-TTS, we use $t\in[1,T]$ to denote the index of sampling point, and $w_{t}$ represents the weight assigned to the current sampling point, determined by the sampler and $s_{t}$ represents the probability of the current sampling point being selected to participate in the model training.
% \begin{equation}
% s_{n}=\frac{c_{n}}{\sum_{i=1}^{N-1}{c_{n}}}
% \end{equation}
	
\paragraph{Uniform sampler} This sampler serves as a baseline for validating other methods, where each point is chosen with equal probability ($c_{n} = 1$).
\paragraph{Linear sampler} The sampling weight varies linearly with the position of the sampling point, defined as $c_{n}=\alpha\cdot n$, with $\alpha = 1 $ in all experiments.
% The parameter $\alpha$ serves as the linear coefficient, consistently set to 1 in our experiments.
\paragraph{Importance sampler (IS)} Following \citealp{nichol2021improved}, we use the IS to assign weights to sampling points. The formulation is given by $c_{n}=(1-\phi)\frac{\sum_{j=1}^{H}{L}(t,j)}{\sum_{i=1}^{N-1}\sum_{j=1}^{H}{L(i,j)}}+\phi$. Here, $\textrm{L} \in\mathbb{R}^{(N-1)\times H} $ represents a matrix recording historical losses for all sampling points, and $H$ denotes the number of historical losses stored for each point (set to 10 in our experiments). The small quantity $\phi$ serves as a balancing factor, adjusting $c_{n}$. This design modulates the probability of current sampling based on historical losses, thereby prioritizing points with greater significance for model training.

\section{Experiments}
\subsection{Data and Preprocessing}
Our experiments are based on CSTR VCTK \citep{6709856}, LJSpeech \citep{ljspeech17}, and LibriSpeech \citep{panayotov2015librispeech} datasets. CSTR VCTK Corpus includes speech data from $110$ English speakers, while LJSpeech features $13,100$ short audio clips, totaling around $24$ hours. For zero-shot experiments, the LibriTTS corpus is used for model training. All samples are resampled to $22,050$ Hz. The test set consists of $512$ randomly selected speech samples, and we assess the model's performance with various objective and subjective metrics. In pre-processing, mel-spectrograms has 80 frequency bins, generated with a window size of $25$ ms and a frameshift of $10$ ms. Ground truth pitch, duration, and energy are computed using the PyWorld toolkit\footnote{\url{https://github.com/JeremyCCHsu/Python-Wrapper-for-World-Vocoder}}.
\subsection{Baseline Models}
\paragraph{Reference and Reference (Voc.)}
Reference denotes the ground truth. The process of obtaining the Reference (voc.) involves transforming the original reference speech into mel-spectrograms, followed by the subsequent reconstruction of speech using HiFi-GAN \citep{kong2020hifi}
\paragraph{FastSpeech2} NAR transformer architecture \citep{ren2019fastspeech}, generating speech in parallel for faster inference. Utilizing mel-spectrogram prediction, duration prediction, and variance modeling, it achieves high efficiency and accuracy in synthesizing speech.
\paragraph{VITS} The VITS model \citep{kim2021conditional} combines variational inference, normalizing flows, and adversarial training. It introduces a stochastic duration predictor to synthesize diverse rhythms, capturing natural variability in speech.
\paragraph{DiffSpeech  \& DiffGAN-TTS}
DiffSpeech \citep{liu2022diffsinger} and DiffGAN-TTS \citep{liu2022diffgan} are diffusion-based TTS architectures. Both architectures focus on addressing real-time speech synthesis in TTS systems, which diffusion models often struggle with due to slow sampling. DiffGAN-TTS addresses the challenge by incorporating additional adversarial training.

\begin{table*}[t]
\centering
\renewcommand\tabcolsep{9pt}
%\small
\scalebox{0.7}
{
\begin{tabular}{lccccccccccc} %{lccccccccccc}
\toprule 
Model& FFE$\downarrow$ & S.Cos$\uparrow$ & mfccFID$\downarrow$ & melFID$\downarrow$ & mfccRecall$\uparrow$ & MCD$\downarrow$ & SSIM$\uparrow$ & mfccCOS$\uparrow$ & F0$\downarrow$ & WER$\downarrow$  \\
\midrule
\model{}(T=1)&\textbf{0.3387}&\textbf{0.8396}&\textbf{39.17}&\textbf{7.58}&0.3946&\textbf{5.91}&\textbf{0.4772}&\textbf{0.7599}&119.29&\textbf{0.0688}\\\midrule
w/o CM&0.3364   &   0.8351  & 43.13   & 10.74  & 0.4010   & 5.98  &   0.4626 &   0.7545  &122.69 & 0.0832 \\
w/o IS &   0.3351  &  0.8333   &56.31  &10.08   & \textbf{0.4015}  &5.98 &  0.4396 & 0.7456 &\textbf{118.87} & 0.0872    \\
\bottomrule
\end{tabular}
}
\caption{Ablation study on VCTK (T=1).}
\label{ablation}
\end{table*}

\begin{table*}[t]
	\centering
	\renewcommand\tabcolsep{6pt}
	%\small
	\scalebox{0.7}
	{
		\begin{tabular}{lccccccccccc}  
			\toprule 
			Simplers&FFE$\downarrow$&S.Cos$\uparrow$&mfccFID$\downarrow$&melFID$\downarrow$&mfccRecall$\uparrow$&MCD$\downarrow$&SSIM$\uparrow$& mfccCOS$\uparrow$ & F0$\downarrow$&WER$\downarrow$&MOS$\uparrow$ \\
			\midrule
			Uniform&\textbf{0.3351}&0.8333&56.31&10.08&0.4015&5.98&0.4396&0.7456&118.87&0.0872&3.8133(±0.0727) \\
			Linear($\scriptstyle\nearrow$)&0.3367&0.8356&63.11&11.35&\textbf{0.4297}&6.03&0.4549&0.7485&\textbf{118.74}&0.0822&3.3278(±0.0803) \\
			Linear($\scriptstyle\searrow$)&0.3403&0.8315&54.58&11.05&0.4102&6.02&0.4694&0.7454&120.32&0.0861&3.5676(±0.1488)\\
			IS&0.3387&\textbf{0.8396}&\textbf{39.17}&\textbf{7.58}&0.3946&\textbf{5.91}&\textbf{0.4772}&\textbf{0.7599}&119.29&\textbf{0.0688}&\textbf{3.9107(±0.1254)} \\
			\bottomrule
		\end{tabular}
	}
	\caption{Performance under different sampler.}
	\label{wieghted simpler}
\end{table*}

\subsection{Model Configuration}
The transformer encoder and the variance adaptor of the \model{} adopt identical network structures and hyper-parameters as those in FastSpeech2. The former is composed of 4 feed-forward transformer (FFT) blocks, where the kernel size and filter size are set to 256, 2, 9, and 1024, respectively. The latter continues to consist of a duration predictor, a pitch predictor, and an energy predictor. The CM-Decoder adopts a structure similar to WaveNet, employing 1D convolution to process the noisy mel spectrogram, followed by activation through the ReLU. Speaker-IDs are activated through WaveNet residual blocks and transformed into embedding vectors. The diffusion step $t$ is encoded using sinusoidal positional encoding as in \citet{song2023consistency}. The mel decoder comprises 4 FFT blocks. The number of parameters in
our model is 28.6 million. 

%%模型基于什么框架
%%模型超参（学习率、模型有多少层层和头CM参数）
%%enconder
%%gpu
%%模型参数
\subsection{Training and Inference}
We conduct all experiments using a single NVIDIA Tesla V100 GPU with $32$ GB. 
The average runtime of training under VCTK, LJSpeech, and LibriSpeech is $34.2$ hours, $42.8$ hours, and $45.6$ hours, respectively. The training employs the multi-speaker dataset VCTK, and speaker embeddings, computed using \citet{li2017deep}, have a dimension of $512$. In our experiments, we randomly select $512$ samples for testing, utilizing the remaining for training. The batch size during training is $32$. We train all the models for $300K$ steps. Following the same learning rate schedule in DiffGAN-TTS, we use an exponential learning rate decay with rate $0.999$ for training and the initial learning rate is $10e^{-4}$. In addition, \citet{song2023consistency} find that periodically adjusting sub-interval $N$ and decay constant $\mu$ in Eq~\ref{decay} during training, following schedule functions $N(k)$ and $\mu(k)$ based on training steps $k$, improves performance. In this paper, we adopts the same strategy as outlined in \citet{song2023consistency}. 
%We performe comparative analyses with FastSpeech2, VITS, DiffSpeech, and DiffGAN-TTS. Specifically, we compared with DiffGAN-TTS at various synthesis steps. Additionally, training on the single-speaker dataset LJSpeech, we conducted a performance comparison with the CoMoSpeech model. To assess zero-shot performance on both VCTK and LJSpeech, we trained on LibriTTS train-clean-100. 
\subsection{Evaluation Metrics}
\paragraph{Objective metrics} 
In our rigorous evaluation of speech synthesis, we leverage a diverse array of objective metrics to holistically appraise the synthesized output's quality and efficiency. This multifaceted set of metrics encompasses the F0 Frame Error (FFE) for evaluating fundamental frequency tracking, Speaker Cosine Similarity (SCS) to gauge the similarity of speaker embeddings, and Fréchet Inception Distance (FID) based on Mel-Frequency Cepstral Coefficients (mfccFID) for a comprehensive assessment of spectrogram divergence. Furthermore, we incorporate metrics such as mfccRecall, MCD24, SSIM, mfccCOS, Word Error Rate (WER), and F0 to provide nuanced insights into various dimensions of synthesis performance.  Detailed descriptions in given in Appendix ~\ref{appendix:metrics}.

\paragraph{Subjective metrics} The Mean Opinion Score (MOS), as introduced in \citet{chu2006objective}, serves as a pivotal metric for evaluating the perceived quality of the synthesized audio. In our evaluation, we involve presenting a carefully curated test set with $30$ samples to $20$ listeners experienced in NLP and speech processing and soliciting their subjective opinions. Participants are then tasked with rating the quality of the synthesized audio on a scale ranging from $1$ to $5$. MOS is a metric that is highly affected by the listeners' subjective judgment. We evaluate the MOS metrics in different tables separately, which causes the MOS of CM-TTS(T=1) to be slightly different rather than identical.
\begin{table*}[t]
\centering
\renewcommand\tabcolsep{6.6pt}
%\small
\scalebox{0.7}
{
\begin{tabular}{lccccccccccc}  
\toprule 
Loss&FFE$\downarrow$&S.Cos$\uparrow$&mfccFID$\downarrow$&melFID$\downarrow$&mfccRecall$\uparrow$&MCD$\downarrow$&SSIM$\uparrow$& mfccCOS$\uparrow$ & F0$\downarrow$&WER$\downarrow$&MOS$\uparrow$ \\
\midrule
$l_{1}$&0.3387&\textbf{0.8396}&39.17&7.5772&0.3946&5.9093&\textbf{0.4772}&0.7599&119.29&\textbf{0.0688}&\textbf{3.9052(±0.0415)}\\
$l_{1}^{\textit{w/o}~padding}$&0.3374&0.8379&43.28&10.16&0.3961&\textbf{5.7815}&0.4593&\textbf{0.7606}&\textbf{117.45}&0.0741&3.8117(±0.1005)\\  \midrule
$l_{2}$&0.3368&0.8320&\textbf{38.73}&\textbf{8.49}&\textbf{0.4062}&5.8836&0.4505&0.7573&120.05&0.0751&3.8726(±0.1971)\\
$l_{2}^{\textit{w/o}~padding}$&\textbf{0.3366}&0.8294&48.09&12.14&0.3841&5.8355&0.4613&0.7585&118.52&0.0756&3.8604(±0.1436)\\
\bottomrule
\end{tabular}
}
\caption{Effect on performance due to padding under different loss. $l_{1}$ and $l_{2}$ represent the loss with padding, whereas $l_{1}^{\textit{w/o}~padding}$ and $l_{2}^{\textit{w/o}~padding}$ represent loss calculation without considering padding.}
\label{Loss-Padding}
\end{table*}

\begin{table*}[h]
	\centering
	\setlength\tabcolsep{3.8pt} 
	%\small
	\scalebox{0.7}
	{
		\begin{tabular}{lccccccccccc} 
			\toprule 
			Model&FFE$\downarrow$&S.Cos$\uparrow$&mfccFID$\downarrow$&melFID$\downarrow$&mfccRecall$\uparrow$&MCD$\downarrow$&SSIM$\uparrow$&mfccCOS$\uparrow$&F0-RMSE$\downarrow$&WER$\downarrow$ &MOS$\uparrow$\\
			\midrule
			DiffGAN-TTS(T=1)&0.4134&0.6874&283.77&44.47&0.1901&9.00&0.2712&0.5351&135.79&0.0488&3.4607(±0.1880)\\
			DiffGAN-TTS(T=2)&\textbf{0.4107}&0.6908&254.84&36.44&0.1950&9.05&0.2764&0.5356&133.96&\textbf{0.0465}&3.5067(±0.1573)\\
			DiffGAN-TTS(T=4)&0.4112&0.6915&256.75&36.50&0.2023&\textbf{9.05}&0.2709&0.5343&135.56&0.0501&3.5893(±0.0298)\\  \midrule
            \model{}(T=1)&0.4219&0.7108&157.91&26.75&0.2072&9.16&0.2829&0.5548&131.27&0.0536&3.8715(±0.0896)\\ 
            \model{}(T=2)&0.4225&0.7107&155.91&\textbf{26.34}&\textbf{0.2135}&9.16&0.2836&\textbf{0.5557}&\textbf{131.13}&0.0536&3.8387(±0.1521)\\
			\model{}(T=4)&0.4226&\textbf{0.7110}&\textbf{155.56}&26.36&0.2089&9.18&\textbf{0.2845}&0.5553&132.04&0.0530&\textbf{3.9221(±0.1016)}\\
			\bottomrule
		\end{tabular}
	}
	\caption{The zero-shot performance of \model{} and DiffGAN-TTS on VCTK for synthesis steps $1$, $2$, and $4$.}
	\label{zeroshot_trained_on_lib_test_on_VCTK}
\end{table*}

\section{Results and Discussion}
\paragraph{Comparison with baselines}
The outcomes of our experiments, comparing the proposed model against various baseline models, are presented in Table~\ref{baselines}. Notably, our model (\model{}) demonstrates a significant performance advantage over Fastspeech2, VITS, and DIffSpeech in objective evaluations. The results also affirm the efficacy of \model{} when pitted against DiffGAN-TTS; the proposed TTS architecture outperforms DiffGAN-TSS across the majority of metrics. Particularly noteworthy is \model{}'s superior performance in single-step generation ($T=1$), where it outperforms DiffGAN-TSS across all objective metrics, with only a minimal gap observed in $f_{0}$. Furthermore, when evaluating speaker similarity (S.Cos), \model{} achieves the highest S.Cos score of $0.8401$, underscoring its effectiveness in multi-speaker speech generation.

% \begin{table*}[h]
% \centering
% \setlength\tabcolsep{8.5pt} 
% \scalebox{0.65}
% {
% \begin{tabular}{ccccccccc} 
% 			\toprule 
%     \multirow{2}{*}{\textbf{VCTK}}&\multicolumn{4}{c}{Pitch}&\multicolumn{4}{c}{Duration}\\ \cline{2-9}
% 	&Mean$\downarrow$&Std$\downarrow$&Skew$\downarrow$&Kurt$\downarrow$&Mean$\downarrow$&Std$\downarrow$&Skew$\downarrow$&Kurt$\downarrow$\\ \midrule
%     DiffGAN-TTS(T=1)&12.95&22.19&\textbf{3.33}&\textbf{15.75}&1.47&0.56&1.52&4.84\\
%     CM-TTS(T=1)&\textbf{12.36}&\textbf{21.53}&3.40&16.37&\textbf{1.36}&\textbf{0.54}&\textbf{1.43}&\textbf{4.83}\\
%     \bottomrule
% \end{tabular}
% }
% \caption{The prosody similarity between synthesized and reference speech in terms of the difference in mean (Mean), standard variation (Std), skewness (Skew), and kurtosis (Kurt) of pitch and duration.}
% \label{pitch_duration_vctk}
% \end{table*}
We conduct a subjective evaluation to compare the naturalness and quality of synthesized speech against a reference sample. The MOS scores from the listening test, showcased in Table~\ref{baselines}, reveal \model{} achieving an impressive MOS of $3.9618$. This marks a substantial advancement over DiffSpeech and a significant outperformance of DiffGAN-TTS in overall performance.
% We compare our model with typical non-autoregressive speech synthesis models, including FastSpeech 2 and DiffSpeech and the one-step synthesis model DiffGAN-TTS. We uniformly set the training steps of all models to 300K to ensure sufficient convergence. For DiffSpeech, we configure the diffusion steps as 1000 and for DiffGAN-TTS and CM-TTS, we configure the synthesis step to be one. As shown in Table~\ref{compare to baselines}, we can observe results in the following aspects: 1) Under the single-step synthesis condition, CM-TTS outperforms DiffGAN-TTS in all other evaluation metrics, except for $FFE$. This strongly demonstrates the effectiveness of the CM architecture. 2) Compared to DiffSpeech based on the traditional diffusion architecture, although CM-TTS did not surpass it in subjective evaluation (MOS), it significantly outperformed it in metrics such as S.Cos, mfccFID, MCD, and approached it closely in the FFE metric. Particularly, in terms of synthesis speed, it achieved a performance that is hundreds of times faster than DiffSpeech. This can be attributed to the effective denoising ability of the consistent framework for noisy spectrograms. 3) Relative to the traditional F, CM-TTS exhibits a $XX\%$ improvement in the COS metric, demonstrating the effectiveness of the CM architecture in learning the distribution of speech data.

\paragraph{Ablation study} To verify the individual contributions of CT and IS to the model's performance, we conduct ablation experiments by separately removing CT and IS, with the synthesis steps set to 1. The experimental results are shown in Table~\ref{ablation}. The results indicate that simultaneous use of both CT and IS samplers leads to notable improvements across multiple metrics, particularly in reducing WER. This underscores their significant contribution to the overall performance of the model. 

\paragraph{Few-step speech generation}
In evaluating single-step synthesis performance, we can observe from Table~\ref{baselines} \model{} that consistently surpasses DiffGAN-TTS across all metrics, with a marginal difference observed in the F0-RMSE. When extending to a multi-step synthesis scenario ($T=4$), \model{} outperforms DiffGAN-TTS in all metrics, except for melFID ($7.34$ compared to $6.58$). These findings emphasize that, beyond its impressive single-step synthesis capabilities, our proposed method demonstrates robust synthesis proficiency in scenarios involving multiple iterative steps.

\paragraph{Length robustness during training}
Incorporating padding in the model's loss calculation is common, especially for variable-length sequences in training. The goal is to guide the model in capturing meaningful representations from both genuine input data and padded segments. TTS models face challenges in handling diverse input texts during training. To assess the model's resilience and investigate the impact of padding, we conduct experiments comparing the inclusion or exclusion of the padding portion in the loss calculation ($\mathcal{L}_{mel}$). Results in Table~\ref{Loss-Padding} demonstrate that including the padding portion improves the overall performance of the model. We experiment with both $l_1$-norm and $l_2$-norm while computing $\mathcal{L}_{mel}$ in Eq.~\ref{equation_recon}.

\begin{table*}[t]
\centering
\renewcommand\tabcolsep{8.6pt}
%\small
\scalebox{0.7}
{
\begin{tabular}{lccccccccccc}  
\toprule 
Model&FFE$\downarrow$&S.Cos$\uparrow$&mfccFID$\downarrow$&melFID$\downarrow$&mfccRecall$\uparrow$&MCD$\downarrow$&SSIM$\uparrow$& mfccCOS$\uparrow$ & F0$\downarrow$&WER$\downarrow$ \\
\midrule
Reference (voc.)&0.1427&0.9424&31.98&3.48&0.5644&4.57&0.8132&0.8457&89.21&0.0412\\\midrule
DiffGAN-TTS(T=2)&0.3411&0.8333&\textbf{38.64}&\textbf{7.79}&0.3974&5.94&\textbf{0.4610}&0.7581&117.19&0.0827\\
with IS &\textbf{0.3397} & \textbf{0.8397}&42.96& 7.92&   \textbf{0.3990}  &\textbf{5.86} &0.4580 & \textbf{0.7582} &\textbf{115.38}&\textbf{0.0720} \\\midrule
DiffGAN-TTS(T=4)&0.3465&0.8358&\textbf{37.11}&\textbf{6.58}&0.3662&5.94&0.4614&0.7571&120.10&0.0751\\
with IS& \textbf{0.3405}&\textbf{0.8403}&43.81 & 7.89 &  \textbf{0.3870 }&\textbf{5.87} &\textbf{0.4641}&  \textbf{ 0.7590} & \textbf{115.89} & \textbf{0.0704}\\
\bottomrule
\end{tabular}
}
\caption{Performance of DiffGAN with and without IS.}
\label{diffgan-LSM}
\end{table*}

\begin{table}[t]
\centering
\setlength\tabcolsep{6pt} 
\scalebox{0.7}
{
\begin{tabular}{llcccc} 
			\toprule 
    % \multirow{2}{*}{\textbf{VCTK}}&\multicolumn{4}{c}{Pitch}&\multicolumn{4}{c}{Duration}\\ \cline{2-9}
	Prosody &Model&Mean$\downarrow$&Std$\downarrow$&Skew$\downarrow$&Kurt$\downarrow$ \\ \midrule %&Mean$\downarrow$&Std$\downarrow$&Skew$\downarrow$&Kurt$\downarrow$\\ \midrule
    % DiffGAN-TTS(T=1)&12.95&22.19&\textbf{3.33}&\textbf{15.75}&1.47&0.56&1.52&4.84\\
    % CM-TTS(T=1)&\textbf{12.36}&\textbf{21.53}&3.40&16.37&\textbf{1.36}&\textbf{0.54}&\textbf{1.43}&\textbf{4.83}\\
    \multirow{2}{*}{Pitch}&DiffGAN-TTS(T=1)&12.95&22.19&\textbf{3.33}&\textbf{15.75} \\
    &\model{}(T=1)&\textbf{12.36}&\textbf{21.53}&3.40&16.37 \\  \midrule 
    \multirow{2}{*}{Duration}&DiffGAN-TTS(T=1)&1.47&0.56&1.52&4.84 \\
    &\model{}(T=1)&\textbf{1.36}&\textbf{0.54}&\textbf{1.43}&\textbf{4.83} \\
    \bottomrule
\end{tabular}
}
\caption{The prosody similarity between synthesized and reference speech of pitch and duration.}
\label{pitch_duration_vctk}
\end{table}

\paragraph{The impact of weighted sampler} 
In this subsection, we conduct experiments to explore the impact of different sampling methods, as discussed in Section~\ref{sec:Weighted Sampler}, on the performance of the \model{}. The results presented in Table~\ref{wieghted simpler} reveal a significant enhancement in the \model{}'s performance across various metrics when the IS sampler is employed. Notably, S.Cos exhibits an improvement to $0.8396$, indicating enhanced speaker similarity with the use of the IS sampler. Furthermore, as illustrated in the Figure~\ref{simpler_cm_loss_convergence}, we observe there is no significant impact on the convergence of \model{} when utilizing a different sampler. To further explore the generalization of IS, we apply it to DiffGAN. The experimental results, as shown in Table~\ref{diffgan-LSM}, strongly demonstrate that IS can bring significant improvements across most metrics.

% \paragraph{Efficiency} \hl{To validate the effectiveness of the different samplers we designed, we set the model training steps to 900K.} This allows for better differentiation of their performance and their impact on model convergence. As presented in Table~\ref{wieghted simpler}, the LSM sampler achieves optimal performance across all six metrics with particularly notable results on melFID. It achieves the second-best performance in FFE, F0, SSIM, and mfccFID. This strongly indicates that the sampling probability does impact the effectiveness of the model. Moreover, from the results, it can be seen that our designed LSM sampler is more reasonable compared to the uniform and linear samplers.
% \paragraph{Convergence}
% As presented in Figure~\ref{simpler_loss}, we can observe that the convergence speed of the Linear($\scriptstyle\nearrow$) sampling method is noticeably slower, while the other three are essentially the same. Hence, it is evident that different sampling methods indeed influence the convergence of model training. Taking into account both efficiency and convergence factors, the LSM sampler demonstrates superior performance. 
\paragraph{Generalization to unseen speakers}
% \hl{Zero-Shot Results}
To assess how well \model{} performs with speakers it hasn't seen before, we train the model on the LibriTTS \citep{zen2019libritts}(train-clean-100) dataset, which mainly contains longer input texts. To test its zero-shot performance, we randomly selected $512$ speech samples from VCTK and LJSpeech datasets. In Table ~\ref{zeroshot_trained_on_lib_test_on_VCTK}, we compare DiffGAN and \model{} on VCTK for different generation steps ($T=1, 2, \& 4$). Additionally, we use an alignment tool to get phoneme-level duration and pitch and compute the prosody similarity between the synthesized and the reference speech. The results are displayed in Table ~\ref{pitch_duration_vctk}. Interestingly, in multi-speaker scenarios, \model{} consistently outperforms the baseline DiffGAN-TTS. However, in single-speaker scenarios (see Table ~\ref{pitch_duration_ljspeech}), DiffGAN-TTS outperforms \model{}. For more details on zero-shot performance on LJSpeech, please refer to Appendix \ref{sec:ZERO-SHOT-LJSpeech}.

% The model is trained on the Libritts(train-clean-100) dataset, which predominantly consists of longer sentences and multi-speakers. Table ~\ref{zeroshot_trained_on_lib_test_on_VCTK} and Table ~\ref{pitch_duration_vctk} are the Zero-shot performance on VCTK, Table ~\ref{zeroshot_trained_on_lib_test_on_LJSpeech} and Table ~\ref{pitch_duration_ljspeech} is the Zero-shot performance on LJSpeech.

\section*{Conclusion}
In this work, we introduced \model{}, a novel architecture focused on real-time speech synthesis. CM-TTS leverages consistency models, steering away from the complexities associated with adversarial training and pre-trained model dependencies. Through comprehensive evaluations, our results underscore the effectiveness of \model{} over established single-step speech synthesis architectures. This marks a significant improvement in promising avenues for applications ranging from voice assistant systems to e-learning platforms and audiobook generation. The future work entails advancing training through the utilization of diverse datasets, thereby enhancing the \model{} to generalize better across previously unseen speakers. 

%\newpage
\section*{Limitations}
% Regarding the model: The proposed CM-TTS framework in the paper mainly optimizes and improves the training mechanism of the model. For the purpose of facilitating comparative experiments, there hasn't been further exploration of the network's inherent structure, such as the number of layers or residual modules. Future work could involve lightweight studies on the network itself, which would better leverage the performance of CM-TTS.
% Regarding the task: The experiments in the paper are focused solely on TTS tasks, without extending to other similar tasks such as sound generation. Future work could involve experimental validation for other related tasks.
In terms of the model, the presented \model{} framework primarily optimizes and enhances the training mechanism, aiming to facilitate comparative experiments. However, the inherent structure of the network, including aspects like the number of layers or residual modules, hasn't been extensively explored for this paper. Future endeavors could delve into lightweight studies focusing on the network itself, potentially enhancing the overall performance of \model{}.

Regarding the task, the experiments conducted in this paper exclusively center around TTS tasks, without extending to other related tasks such as sound generation. Future work could encompass experimental validation across a broader spectrum of tasks, providing a more comprehensive assessment.
\section*{Ethics Statement}
Given the ability of CM-TTS to synthesize speech while preserving the speaker's identity, potential risks of misuse, such as deceiving voice recognition systems or impersonating specific individuals, may arise. In our experiments, we operate under the assumption that users willingly agree to be the designated speaker for speech synthesis. In the event of the model's application to unknown speakers in real-world scenarios, it is imperative to establish a protocol ensuring explicit consent from speakers for the utilization of their voices. Additionally, implementing a synthetic speech detection model is recommended to mitigate the potential for misuse.
\section*{Acknowledgements}
We thank the anonymous reviewers for their constructive feedback. This work was supported in part by the National Key Research and Development Program of China under grant 2022YFF0902701, the National Natural Science Foundation of China under grant U21A20468, 61921003, U22A201339, the Fundamental Research Funds for the Central Universities under Grant 2020XD-A07-1, and the BUPT Excellent Ph.D. Students Foundation under Grant CX2023224.

% Additional elements were taken from the formatting instructions of the \emph{International Joint Conference on Artificial Intelligence} and the \emph{Conference on Computer Vision and Pattern Recognition}.

% Entries for the entire Anthology, followed by custom entries
% \newpage
\bibliography{custom}
\newpage
\appendix
\clearpage\newpage
\onecolumn 
\section{Experiments on LJSpeech}
\label{sec:LJSpeech}
% \paragraph{Comparison to Baselines}
% We also tested the model's performance on various metrics using the single speaker dataset LJSpeech.The results are shown in Table~\ref{app_LJSpeech_baseline}. Similarly, we set the model training steps to 300K and the speech synthesis steps to one. It can be seen that our CM-TTS demonstrates similarly impressive advantages in one-step synthesis on single speaker datasets. 
% We also thoroughly compare the performance of CM-TTS and DiffGAN-TTS on the single-speaker dataset LJSpeech. The results are shown in Table~\ref{app_LJSpeech_baseline}. It can be seen that when the model training steps are 300K and the synthesis steps are 4, our CM-TTS achieves a score of 0.9010 on the S.Cos metric and a score of 2.97 on the melFID metric. Both scores represent the optimal performance in all training and synthesis scenarios. Moreover, when the model training steps are 100K and the synthesis steps are one, our model outperforms DiffGAN-TTS in multiple metrics. This clearly demonstrates that our model also performs very well on single speaker scenarios.

Our \model{} model, trained for $300$K steps on the LJSpeech single speaker dataset, exhibits impressive performance in $1$, $2$, and $4$-step synthesis, detailed in Table~\ref{app_LJSpeech_baseline}. Compared to DiffGAN-TTS, \model{} achieves optimal scores (S.Cos: 0.9010, melFID: 2.97) across varied training and synthesis scenarios, highlighting its effectiveness in single-speaker scenarios.

% To further compare the performance of CM-TTS and DiffGAN-TTS, we examine the convergence of these two models at different training steps.As presented in Figure~\ref{training_loss}, At the beginning of the model training, the convergence of the two is relatively consistent. However, as the number of model training steps increases, CM-TTS exhibits a noticeably better convergence, indicating superior fitting performance compared to DiffGAN-TTS.

\begin{figure}[t!]
	%\flushleft
	\centering
	\includegraphics[width=0.50\linewidth]{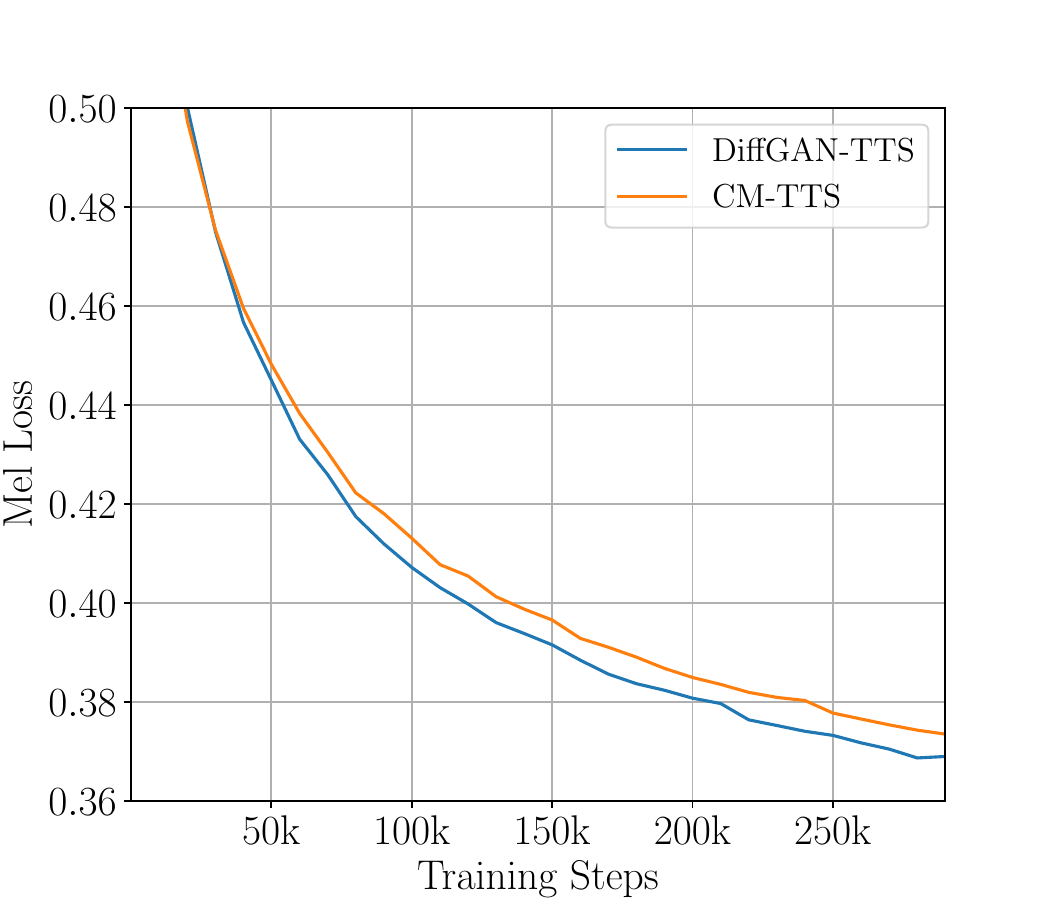}
	\caption{ An Illustration of the Convergence of Loss Across DiffGAN-TTS and CM-TTS.}
	\label{training_loss} 
\end{figure}

In a detailed performance comparison between \model{} and DiffGAN-TTS, we analyze the convergence of these models across various training steps, as illustrated in Figure~\ref{training_loss}. Initially, both models exhibit relatively consistent convergence. However, as the training steps increase, \model{} demonstrates significantly better convergence, indicating superior fitting performance when compared to DiffGAN-TTS.

\begin{table*}[h]
	\centering
	\renewcommand\tabcolsep{2pt}
	%\small
	\scalebox{0.7}
	{
		\begin{tabular}{lcccccccccccc}  
			\toprule Model&FFE$\downarrow$&S.Cos$\uparrow$&mfccFID$\downarrow$&melFID$\downarrow$&mfccRecall$\uparrow$&MCD$\downarrow$&SSIM$\uparrow$& mfccCOS$\uparrow$ & F0$\downarrow$ & RTF$\downarrow$ & WER$\downarrow$ &MOS$\uparrow$ \\
			\midrule
			Reference &-&-&-&4.49e-11&0.7013&-&-&-& -&-&0.0808&- \\
			Reference (voc.)&0.0891&0.9861&0.8323&0.11&0.6768&3.1995&0.9310&0.9589&67.61&-&0.0712&4.8667(±0.0315) \\\midrule
			FastSpeech2&0.4877&0.8825&36.31&5.28&0.2121&6.1157&0.6468&0.7985&135.26&-&0.0944&3.5742(±0.2309)\\
			DiffSpeech&0.4885&0.8742&27.45&4.38&0.2775&7.0267&0.5562&0.7332&132.59&- &0.1171&3.1668(±0.1378)\\
			CoMoSpeech&0.4900&0.8666&369.96&17.81&0.2865 &7.7416&0.5660&0.7275&144.23&- &0.0823&3.5583(±0.2421)\\
			VITS&0.4820&0.8811&264.89&17.82&0.3192 &7.0700&0.6248&0.7776&\textbf{123.24}&- &0.0847&3.6234(±0.0252)\\ \midrule
			DiffGAN-TTS(T=1)&0.4872&0.8959&27.22&3.70&0.2527&\textbf{6.0798}&0.6530&0.7991&136.80&-&0.0697&3.7142(±0.1390)\\
			DiffGAN-TTS(T=2)&\textbf{0.4818}&0.8995&25.03&3.09&\textbf{0.2463}&6.1205&0.6547&\textbf{0.7995}&133.71&-&0.0749&3.6813(±0.0561)\\
			DiffGAN-TTS(T=4)&0.4856&0.8969&\textbf{23.48}&3.15&0.2590&6.0856&0.6539&0.7991&136.50&-&\textbf{0.0693}&3.7258(±0.0087)\\ \midrule
            \model{}(T=1)&0.4860&0.9009&24.52&2.97&0.2586&6.0978&\textbf{0.6558}&0.7989&135.58&-&0.0727&3.8353(±0.0179)\\
            \model{}(T=2)&0.4861&0.9010&24.70&2.97&0.2597&6.0978&0.6553&0.7990&136.02&-&0.0725&\textbf{3.7917(±0.1356)}\\
			\model{}(T=4)&04861&$\mathbf{0.9010}$&24.72&$\mathbf{2.97}$&0.2591&6.0965&0.6553&0.7989&136.26&-&0.0725&3.7602(±0.1327)\\
			\bottomrule
		\end{tabular}
	}
	\caption{Objective evaluation: Comparison with baselines on LJSpeech dataset.}
	\label{app_LJSpeech_baseline}
\end{table*}

\section{Zero-shot Performance on LJSpeech}\label{sec:ZERO-SHOT-LJSpeech}
We trained \model{} on the LibriTTS' train-clean-100 dataset and evaluated LJSpeech's zero-shot performance. The results are presented in Table~\ref{zeroshot_trained_on_lib_test_on_LJSpeech} and Table~\ref{pitch_duration_ljspeech}. It is evident that \model{} consistently outperforms in most metrics.

\begin{table*}[h]
\centering
\setlength\tabcolsep{9pt} 
\scalebox{0.7}
{
\begin{tabular}{ccccccccc} 
	\toprule 
    \multirow{2}{*}{\textbf{LJSpeech}}&\multicolumn{4}{c}{Pitch}&\multicolumn{4}{c}{Duration}\\ \cline{2-9}
	&Mean$\downarrow$&Std$\downarrow$&Skew$\downarrow$&Kurt$\downarrow$&Mean$\downarrow$&Std$\downarrow$&Skew$\downarrow$&Kurt$\downarrow$\\ \midrule
    DiffGAN-TTS(T=1)&20.56&32.11&\textbf{3.45}&\textbf{18.34}&\textbf{0.93}&\textbf{0.65}&\textbf{0.75}&4.39\\
    \model{}(1)&\textbf{18.34}&\textbf{29.99}&3.73&21.35&1.08&0.92&1.70&\textbf{4.38}\\
    \bottomrule
\end{tabular}
}
\caption{The prosody similarity between synthesized and prompt speech in terms of the difference in mean (Mean), standard variation (Std), skewness (Skew), and kurtosis (Kurt) of pitch and duration on LJSpeech. \textbf{Best} numbers are highlighted in each column.}
\label{pitch_duration_ljspeech}
\end{table*}

\begin{table*}[h]
	\centering
	\setlength\tabcolsep{3.6pt} 
	%\small
	\scalebox{0.7}
	{
		\begin{tabular}{lcccccccccccc} 
			\toprule 
			Model&FFE$\downarrow$&S.Cos$\uparrow$&mfccFID$\downarrow$&melFID$\downarrow$&mfccRecall$\uparrow$&MCD$\downarrow$&SSIM$\uparrow$&mfccCOS$\uparrow$&F0-RMSE$\downarrow$&WER$\downarrow$&MOS$\uparrow$\\
			\midrule
			DiffGAN-TTS(T=1)&0.5164&0.7278&162.90&21.83&0.2523&8.3634&\textbf{0.4491}&0.6513&170.26&0.1118&3.6047(0.1015±)\\
			DiffGAN-TTS(T=2)&0.5151&\textbf{0.7339}&93.96&13.50&0.2772&8.2702&0.4479&0.6561&164.80&0.1146&3.6212(±0.0771)\\
			DiffGAN-TTS(T=4)&0.5153&0.7315&95.08&13.38&0.2859&\textbf{8.2692}&0.4447&0.6547&161.62&\textbf{0.1094}&3.7361(±0.1802)\\
            \midrule
            \model{}(T=1)&\textbf{0.4934}&0.7271&\textbf{86.90}&10.84&\textbf{0.4013}&8.6616&0.4433&0.6540&148.04&0.1194&3.7205(±0.1097)\\ 
            \model{}(T=2)&0.5060&0.7290&105.34&9.12&0.3082&8.5547&0.4458&0.6587&148.83&0.1190&3.6817(±0.1328)\\
			\model{}(T=4)&0.5081&0.7301&102.35&\textbf{8.91}&0.2876&8.6102&0.4392&\textbf{0.6596}&\textbf{147.38}&0.1264&3.7113(±0.1022)\\
			\bottomrule
		\end{tabular}
	}
	\caption{The zero-shot performance of \model{} and DiffGAN-TTS on LJSpeech. $T$ equal to $1$, $2$ \& $4$ represents steps for synthesis. \textbf{Best} numbers are highlighted in each column.}
	\label{zeroshot_trained_on_lib_test_on_LJSpeech}
\end{table*}

\begin{figure}[t!]
	%\flushleft
	\centering	 
    \includegraphics[scale=0.4]{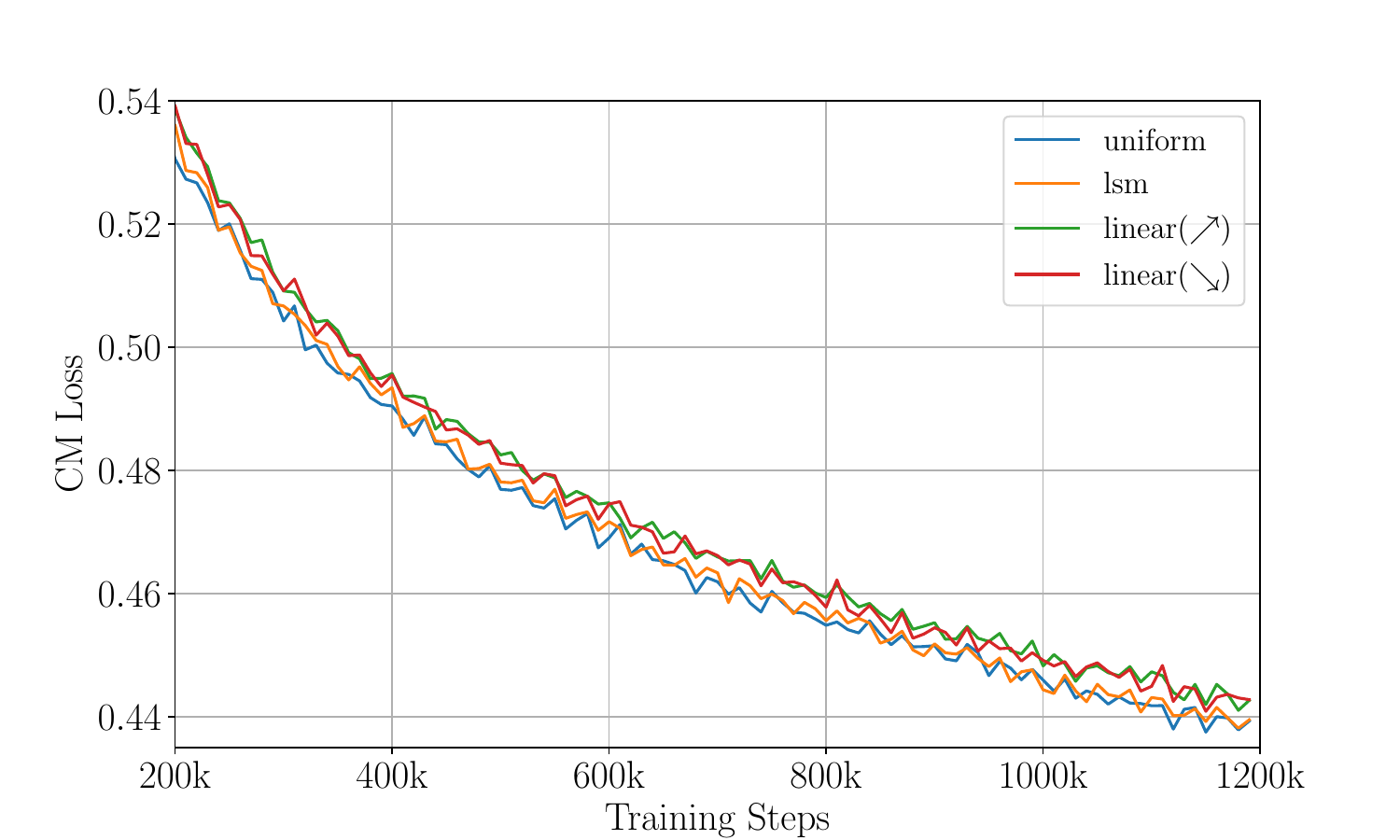}
	\caption{Convergence of loss across different Samplers.}
	\label{simpler_cm_loss_convergence} 
\end{figure}

\section{50 Particularly Hard Sentences}
To evaluate the robustness of \model{}, we follow the practice in \citep{ren2020fastspeech,ping2018deep} and generate $50$ sentences which
are particularly hard for the TTS system. Subjectively assessing the results, we observed that, aside from occasional inaccuracies in pronouncing individual words, the synthesis quality across the majority of examples is notably clear. This observation strongly supports the claim that \model{} exhibits considerable robustness in handling a wide range of linguistic complexities. The specific textual representations for all the sentences are provided below for reference.
% To comprehensively evaluate the robustness of CM-TTS, we meticulously crafted a set of 50 challenging sentences aimed at scrutinizing the model's capabilities in the realm of speech synthesis. This curated selection includes diverse linguistic challenges, such as individual letters, multiple consecutive repeated letters or numbers, sentences with various pause characters ($_$, comma, period), combinations of single letters and words, and sentences featuring special characters like $+$, $()$, $?$, and more. The corresponding synthesized examples for these challenging sentences have been thoughtfully showcased on a dedicated webpage. Subjectively assessing the results, we observed that, aside from occasional inaccuracies in pronouncing individual words, the synthesis quality across the majority of examples is notably clear. This observation strongly supports the claim that our model exhibits considerable robustness in handling a wide range of linguistic complexities. The specific textual representations for all the sentences are provided below for reference.
\begin{itemize}
	\setlength{\itemsep}{0pt}
	\setlength{\parsep}{0pt}
	\setlength{\parskip}{0pt}
	\item[01.] a
	\item[02.] b
	\item[03.] c
	\item[04.] H
	\item[05.] I
	\item[06.] J
	\item[07.] K
	\item[08.] L
	\item[09.] 22222222 hello 22222222
	\item[10.] S D S D Pass zero - zero Fail - zero to zero - zero - zero Cancelled - fifty nine to three - two -
	sixty four Total - fifty nine to three - two -
	\item[11.] S D S D Pass - zero - zero - zero - zero Fail - zero - zero - zero - zero Cancelled - four hundred
	and sixteen - seventy six -
	\item[12.] zero - one - one - two Cancelled - zero - zero - zero - zero Total - two hundred and eighty six -
	nineteen - seven -
	\item[13.] forty one to five three hundred and eleven Fail - one - one to zero two Cancelled - zero - zero to
	zero zero Total -
	\item[14.] zero zero one , MS03 - zero twenty five , MS03 - zero thirty two , MS03 - zero thirty nine ,
	\item[15.] 1b204928 zero zero zero zero zero zero zero zero zero zero zero zero zero zero one seven ole32
	\item[16.] zero zero zero zero zero zero zero zero two seven nine eight F three forty zero zero zero zero zero
	six four two eight zero one eight
	\item[17.] c five eight zero three three nine a zero bf eight FALSE zero zero zero bba3add2 - c229 - 4cdb -
	\item[18.] Calendaring agent failed with error code 0x80070005 while saving appointment .
	\item[19.] Exit process - break ld - Load module - output ud - Unload module - ignore ser - System error -
	ignore ibp - Initial breakpoint -
	\item[20.] Common DB connectors include the DB - nine , DB - fifteen , DB - nineteen , DB - twenty five ,
	DB - thirty seven , and DB - fifty connectors .
	\item[21.] To deliver interfaces that are significantly better suited to create and process RFC eight twenty
	one , RFC eight twenty two , RFC nine seventy seven , and MIME content .
	\item[22.] int1 , int2 , int3 , int4 , int5 , int6 , int7 , int8 , int9 ,
	\item[23.] seven \_\ ctl00 ctl04 ctl01 ctl00 ctl00
	\item[24.] Http0XX , Http1XX , Http2XX , Http3XX ,
	\item[25.] config file must contain A , B , C , D , E , F , and G .
	\item[26.] mondo - debug mondo - ship motif - debug motif - ship sts - debug sts - ship Comparing local
	files to checkpoint files ...
	\item[27.] Rusbvts . dll Dsaccessbvts . dll Exchmembvt . dll Draino . dll Im trying to deploy a new topology
	, and I keep getting this error .
	\item[28.] You can call me directly at four two five seven zero three seven three four four or my cell four two
	five four four four seven four seven four or send me a meeting request with all the appropriate
	information .
	\item[29.] Failed zero point zero zero percent < one zero zero one zero zero zero zero Internal . Exchange .
	ContentFilter . BVT ContentFilter . BVT\_\ log . xml Error ! Filename not specified .
	\item[30.] C colon backslash o one two f c p a r t y backslash d e v one two backslash oasys backslash
	legacy backslash web backslash HELP
	\item[31.] src backslash mapi backslash t n e f d e c dot c dot o l d backslash backslash m o z a r t f one
	backslash e x five
	\item[32.] copy backslash backslash j o h n f a n four backslash scratch backslash M i c r o s o f t dot S h a r
	e P o i n t dot
	\item[33.] Take a look at h t t p colon slash slash w w w dot granite dot a b dot c a slash access slash email
	dot
	\item[34.] backslash bin backslash premium backslash forms backslash r e g i o n a l o p t i o n s dot a s p x
	dot c s Raj , DJ ,
	\item[35.] Anuraag backslash backslash r a d u r five backslash d e b u g dot one eight zero nine underscore
	P R two h dot s t s contains
	\item[36.] p l a t f o r m right bracket backslash left bracket f l a v o r right bracket backslash s e t u p dot e x
	e
	\item[37.] backslash x eight six backslash Ship backslash zero backslash A d d r e s s B o o k dot C o n t a c
	t s A d d r e s
	\item[38.] Mine is here backslash backslash g a b e h a l l hyphen m o t h r a backslash S v r underscore O f
	f i c e s v r
	\item[39.] h t t p colon slash slash teams slash sites slash T A G slash default dot aspx As always , any
	feedback , comments ,
	\item[40.] two thousand and five h t t p colon slash slash news dot com dot com slash i slash n e slash f d
	slash two zero zero three slash f d
	\item[41.] backslash i n t e r n a l dot e x c h a n g e dot m a n a g e m e n t dot s y s t e m m a n a g e
	\item[42.] I think Rich’s post highlights that we could have been more strategic about how the sum total of
	XBOX three hundred and sixtys were distributed .
	\item[43.] 64X64 , 8K , one hundred and eighty four ASSEMBLY , DIGITAL VIDEO DISK DRIVE ,
	INTERNAL , 8X ,
	\item[44.] So we are back to Extended MAPI and C++ because . Extended MAPI does not have a dual
	interface VB or VB .Net can read .
	\item[45.] Thanks , Borge Trongmo Hi gurus , Could you help us E2K ASP guys with the following issue ?
	\item[46.] Thanks J RGR Are you using the LDDM driver for this system or the in the build XDDM driver ?
	\item[47.]  Btw , you might remember me from our discussion about OWA automation and OWA readiness
	day a year ago .
	\item[48.] empidtool . exe creates HKEY\_\ CURRENT\_\ USER Software Microsoft Office Common
	QMPersNum in the registry , queries AD , and the populate the registry with MS employment ID
	if available else an error code is logged .
	\item[49.] Thursday, via a joint press release and Microsoft AI Blog, we will announce Microsoft’s
	continued partnership with Shell leveraging cloud, AI, and collaboration technology to drive
	industry innovation and transformation.
	\item[50.] Actress Fan Bingbing attends the screening of ’Ash Is Purest White (Jiang Hu Er Nv)’ during the
	71st annual Cannes Film Festival
\end{itemize}
\section{Metrics}
\label{appendix:metrics}
We employ 12 metrics to assess the quality and efficiency of speech synthesis. This includes 11 objective metrics and one subjective metric. The following provides a detailed analysis of the calculation methods and objectivity for all the metrics involved in the experiments.
\begin{itemize}

\item \textbf{FFE (Fundamental Frequency Frame Error):}
  \begin{itemize}
  \item FFE, or F0 Frame Error \citep{4960497}, combines Gross Pitch Error (GPE) and Voicing Decision Error (VDE) to objectively evaluate fundamental frequency (F0) tracking methods.
  \item The Fundamental Frequency Frame Error (FFE) quantifies errors during the estimation of the fundamental frequency using the formula:
  \[
  FFE = \frac{1}{N} \sum_{i=1}^{N} \left| F_{0i,\text{estimated}} - F_{0i,\text{actual}} \right|
  \]
  where \(N\) is the total number of frames, \(F_{0i,\text{estimated}}\) is the estimated fundamental frequency of the \(i\)-th frame, and \(F_{0i,\text{actual}}\) is the actual fundamental frequency of the \(i\)-th frame.
  \end{itemize}

\item \textbf{S.Cos (Speaker Cosine Similarity):}
  \begin{itemize}
  \item S.Cos, or Speaker Cosine Similarity, measures the degree of similarity between speaker embeddings corresponding to synthesized speech and ground truth.
  \item The Cosine Similarity is calculated as:
  \[
  \text{Cosine Similarity}(\mathbf{P}, \mathbf{A}) = \frac{\mathbf{P} \cdot \mathbf{A}}{\|\mathbf{P}\| \|\mathbf{A}\|}
  \]
  where \(\mathbf{P} \cdot \mathbf{A}\) is the dot product between speaker embeddings, and \(\|\mathbf{P}\| \|\mathbf{A}\|\) is their Euclidean norm.
  \end{itemize}

\item \textbf{mfccFID (Fréchet Inception Distance based on MFCC):}
  \begin{itemize}
  \item mfccFID calculates the Fréchet Inception Distance (FID) between MFCC features extracted from predicted and actual speech, measuring similarity between their distributions.
  \item The FID formula is given by:
  \[
  FID = \|\mu_p - \mu_a\|^2 + \text{Tr}(\Sigma_p + \Sigma_a - 2(\Sigma_p \Sigma_a)^{1/2})
  \]
  where \(\mu_p\) and \(\mu_a\) are mean vectors, and \(\Sigma_p + \Sigma_a\) is the covariance matrix.
  \end{itemize}

\item \textbf{melFID (Fréchet Inception Distance based on Mel Spectrogram):}
  \begin{itemize}
  \item melFID directly calculates FID between Mel spectrograms of predicted and actual frames.
  \end{itemize}

\item \textbf{mfccRecall:}
  \begin{itemize}
%   \item As described in \citet{kynkaanniemi2019improved},  We denote feature vectors of the real and generated mel spectrograms by $\phi_{r}$ and $\phi_{g}$ respectively, we adopted the MFCC features of the speeches and the corresponding sets of feature vectors by $\Phi_{r}$ and $\Phi_{g}$. We take an equal number of samples from each distribution. Recall is calculated by querying for each
% real image whether the image is within estimated manifold of generated images.
\item As outlined in \citet{kynkaanniemi2019improved}, we denote the feature vectors of real and generated mel spectrograms as $\phi_{r}$ and $\phi_{g}$, respectively. In our approach, we utilized the MFCC features of the speeches, representing the sets of feature vectors as $\Phi_{r}$ and $\Phi_{g}$. We ensured an equal number of samples were drawn from each distribution. Recall is computed by querying, for each real image, whether the image falls within the estimated manifold of generated images.
  \item The formula is:
  	\begin{equation*}
	 	recall(\Phi_{r} , \Phi_{g})=\frac{1}{\left| \Phi_{r} \right|}\sum_{\phi_{r}\in \Phi_{r}}{f(\phi_{r}, \Phi_{g})}		
	\end{equation*} 
 $ f(\phi, \Phi_{g})$ provides a way
	  to determine whether it could be reproduced by the generator.
  \end{itemize}

\item \textbf{MCD (Mel Cepstral Distortion):}
  \begin{itemize}
  \item MCD measures the difference between two acoustic signals in the domain of Mel Cepstral Coefficients (MFCC).
  \item The formula is:
  \[
  MCD = \frac{1}{T} \sum_{t=1}^{T} d(c(p), c(a))
  \]
  where \(T\) is the total number of frames, and \(c(p)\) and \(c(a)\) are the MFCC vectors of real and synthesized speech.
  \end{itemize}

\item \textbf{SSIM (Structural Similarity Index):}
  \begin{itemize}
  \item SSIM measures the similarity between two spectrograms using luminance, contrast, and structure information.
  \item The SSIM formula is given by:
  \[
  \text{SSIM}(p, a) = \frac{(2\mu_p\mu_a + c_1)(2\sigma_{pa} + c_2)}{(\mu_p^2 + \mu_a^2 + c_1)(\sigma_p^2 + \sigma_a^2 + c_2)}
  \]
  where \(p\) and \(a\) are the spectrograms, and \(\mu_p\), \(\mu_a\), \(\sigma_p^2\), \(\sigma_a^2\), \(\sigma_{pa}\), \(c_1\), and \(c_2\) are constants.
  \end{itemize}

\item \textbf{mfccCOS (MFCC Cosine Similarity):}
  \begin{itemize}
  \item mfccCOS measures the similarity between MFCC features of real and predicted speech using the same calculation method as S.Cos.
  \end{itemize}

\item \textbf{F0-RMSE (F0 Root Mean Squared Error):}
  \begin{itemize}
  \item F0-RMSE is a metric measuring the difference between two pitch sequences (fundamental frequency).
  \item The RMSE formula is:
  \[
  \text{RMSE} = \sqrt{\frac{1}{N} \sum_{i=1}^{N} (f_{0,i} - \hat{f}_{0,i})^2}
  \]
  where \(N\) is the total number of frames, \(f_{0,i}\) is the fundamental frequency of the \(i\)-th frame in the real pitch sequence, and \(\hat{f}_{0,i}\) is the fundamental frequency of the \(i\)-th frame in the predicted pitch sequence.
  \end{itemize}

\item \textbf{RTF (Real-time Factor):}
  \begin{itemize}
  \item RTF represents the time (in seconds) required for the system to synthesize one second of waveform.
  \end{itemize}

\item \textbf{MOS (Mean Opinion Score):}
  \begin{itemize}
  \item MOS is an objective evaluation metric obtained through subjective experiments, assessing the quality of speech synthesis.
  \item The MOS formula is:
  \[
  \text{MOS} = \frac{1}{N} \sum_{i=1}^{N} a_i
  \]
  where \(N\) is the number of participants, and \(a_i\) is the score provided by the \(i\)-th participant.
  \end{itemize}

\item \textbf{WER (Word Error Rate):}
  \begin{itemize}
  \item WER measures the disparity between the transcribed text of the model's predicted speech and the actual speech. The calculation of WER includes three types of errors : Insertions, Deletions, and Substitutions.
  \item The WER formula is:
  \[
  \text{WER} = \frac{S+D+I}{N} \times 100
  \]
  where \(S\) is the number of substitution errors, \(D\) is the number of deletion errors, \(I\) is the number of insertion errors and \(N\) is is the total number of words in the transcribed text.
  \end{itemize}
\end{itemize}

\section{Metric}

\begin{figure}[h]
	%\flushleft
	\centering
	\includegraphics[width=0.50\linewidth]{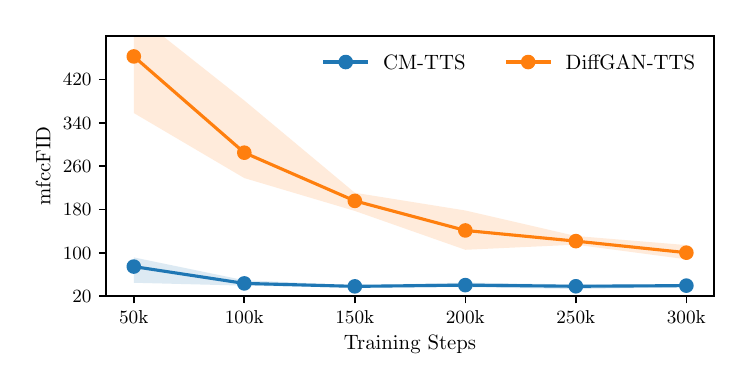}
	\caption{ The trend of DiffGAN-TTS and \model{} on the mfcc-FID metric during training on VCTK.}
	\label{mfcc-FID} 
\end{figure}

\begin{figure}[h]
	%\flushleft
	\centering
	\includegraphics[width=0.50\linewidth]{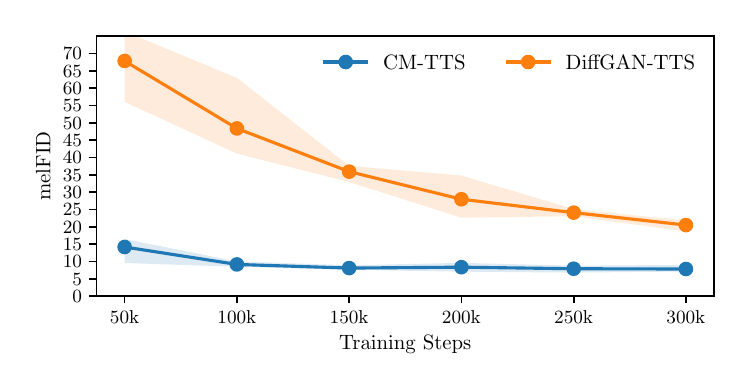}
	\caption{ The trend of DiffGAN-TTS and \model{} on the mel-FID metric during training on VCTK.}
	\label{mel-FID} 
\end{figure}

\begin{figure}[h]
	%\flushleft
	\centering
	\includegraphics[width=0.80\linewidth]{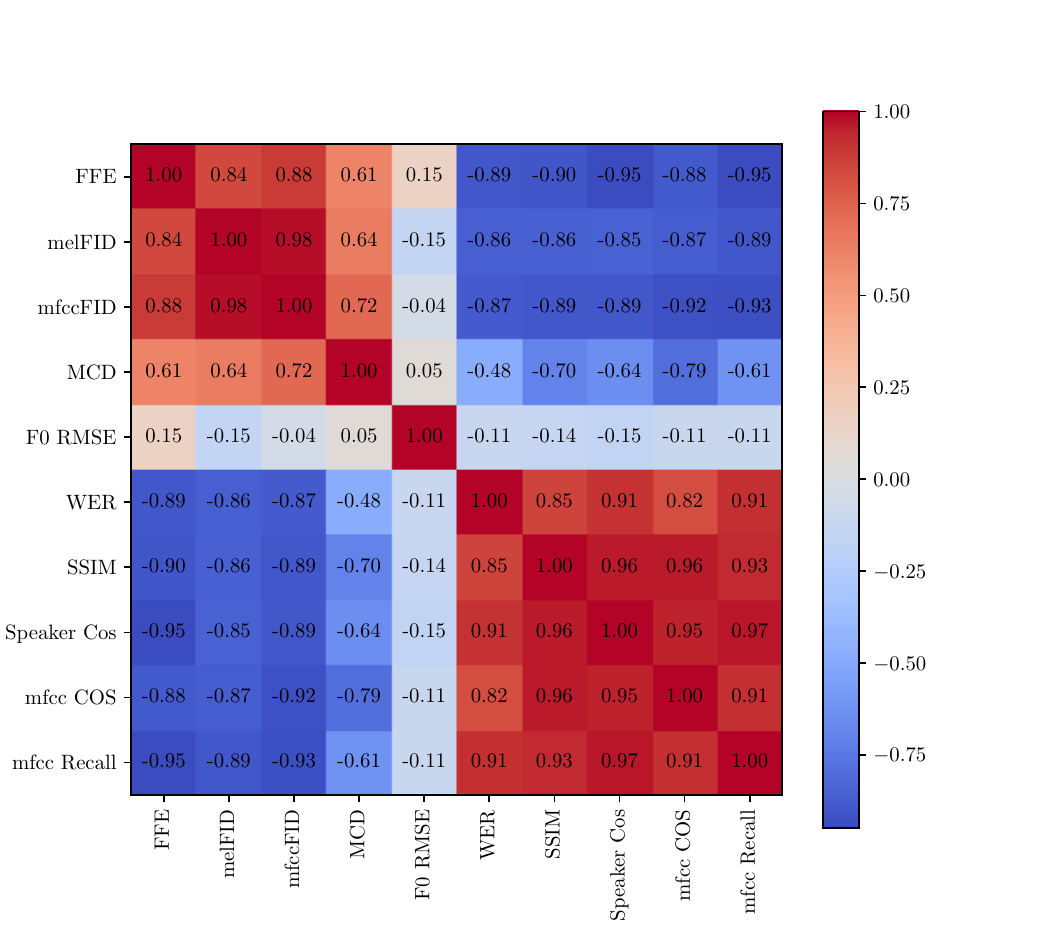}
	\caption{ The Pearson correlation coefficient between different objective evaluation metrics.}
	\label{metric correlation} 
\end{figure}

% From the trend of changes in the metrics, it can be observed that CM exhibits faster convergence and more stable model performance.

As depicted in Figure~\ref{mfcc-FID} and Figure~\ref{mel-FID}, the trend in metric changes highlights that \model{} displays faster convergence and a more stable model performance.

We also explored relationships between various evaluation metrics, calculating trends' similarity using the Pearson coefficient and visualizing the results in Figure~\ref{metric correlation}. Notably, significant correlations were observed among SSIM, Speaker Cos, mfccCOS, and mfcc Recall, indicating closely aligned trends. A strong correlation was also identified between the two types of FID. Conversely, MCD showed a weak relationship with metrics that perform better when lower. F0 RMSE displayed weak correlations with all other metrics, and FFE had a relatively modest relationship with metrics that are optimal when smaller. This study provides valuable insights for speech synthesis quality evaluation, suggesting that when testing only a few metrics, it's advisable to select those with lower correlations, as illustrated in the Figure~\ref{metric correlation}, as evaluation indicators.

\end{document}